\documentclass[preprint,onecolumn,amsmath]{revtex4}

\usepackage{graphicx}
\usepackage{amssymb,amsfonts,amsmath}

\usepackage[utf8]{inputenc}
\usepackage[russian,english]{babel}

\def\s{\sigma}

\def \g{\gamma}    \def \a{\alpha}
\def \w{\omega}    \def \b{\beta} 
\def \s{\sigma}      
\def \e{\epsilon}    
   \def \d{\delta} 
    \def \l{\lambda}

\def \del{\partial}    
 
\def \hf{\tfrac{1}{2}} 
\def \HF{\dfrac{1}{2}}  

\def \>{\rangle} 
\def \<{\langle} 
 
\def\be{\begin{equation}} 
\def\ee{\end{equation}} 
\def\longrightharpoonup{\relbar\joinrel\rightharpoonup}
\def\longleftharpoondown{\leftharpoondown\joinrel\relbar}

\def\longrightleftharpoons{
  \mathop{
    \vcenter{
      \hbox{
      \ooalign{
        \raise1pt\hbox{$\longrightharpoonup\joinrel$}\crcr
	  \lower1pt\hbox{$\longleftharpoondown\joinrel$}
	  }
      }
    }
  }
}

\newcommand \bea {\begin{eqnarray}} 
\newcommand \eea {\end{eqnarray}}

\begin{document}

\title{Fast Bayesian Feature Selection for High Dimensional Linear Regression in Genomics via the Ising Approximation}

\author{Charles K. Fisher}
\affiliation{Deptartment of Physics, Boston University, Boston, MA 02215}

\author{Pankaj Mehta}
\affiliation{Deptartment of Physics, Boston University, Boston, MA 02215}

\begin{abstract}

Feature selection, identifying a subset of variables that are relevant for predicting a response, is an important and challenging component of many methods in statistics and machine learning. Feature selection is especially difficult  and computationally intensive when the number of variables approaches or exceeds the number of samples, as is often the case for many genomic datasets. Here, we introduce a new approach --  the Bayesian Ising Approximation (BIA)  -- to rapidly calculate posterior probabilities for feature relevance in L2 penalized linear regression. In the regime where the regression problem is strongly regularized by the prior, we show that computing the marginal posterior probabilities for features  is equivalent to computing the magnetizations of an Ising model. Using a mean field approximation, we show it is possible to rapidly compute the feature selection path described by the posterior probabilities as a function of the L2 penalty. We present simulations and analytical results illustrating the accuracy of the BIA  on some simple regression problems. Finally, we demonstrate the applicability of the BIA to high dimensional regression by analyzing a gene expression dataset with nearly 30,000 features.

 \end{abstract}

\maketitle

\section{Introduction}

Linear regression is one of the most broadly and frequently used statistical tools. Despite hundreds of years of research on the subject \cite{legendre1805nouvelles}, modern applications of linear regression to large datasets present a number of new challenges. Modern applications of linear regression, such as Genome Wide Association Studies (GWAS), often consider datasets that have at least as many potential variables (or features) as there are data points \cite{mccarthy2008genome}.  Applying linear regression to high dimensional datasets often involves selecting a subset of relevant features, a problem known as feature selection in the literature on statistics and machine learning \cite{guyon2003introduction}. Even for classical least-squares linear regression, it turns out that the associated feature selection problem is quite difficult  \cite{huo2007stepwise}. 

The difficulties associated with feature selection are especially pronounced in genomics and GWAS. In general, the goal  of many genomics studies is to identify a relationship between a small number of  genes and a phenotype of interest, such as height or body mass index  \cite{mccarthy2008genome, peng2010regularized,burton2007genome,subramanian2005gene,wu2009genome}. For example, many GWAS seek to identify specific genetic mutations (called single nucleotide polymorphisms or SNPs) that best explain the variation of a quantitative trait, such as height or body mass index, in a population \cite{yang2012conditional}. Using various techniques, the trait is regressed against binary variables representing the presence or absence of the SNPs in order to find a subset of SNPs that are highly explanatory for the trait \cite{wu2009genome,peng2010regularized}. Although the number of individuals genotyped in such a study may be in the thousands or even tens of thousands, this pales in comparision to the number of potential SNPs which can be in the millions \cite{mccarthy2008genome}. Moreover, the presence or absence of various SNPs tend to be correlated due to chromosome structure and genetic processes that induce so-called linkage disequilibrium  \cite{yang2012conditional}. As a result, selecting the best subset of SNPs for the regression involves a search for the global minimum of a landscape that is both high dimensional (due to the large number of SNPs) and rugged (due to correlations between SNPs). 

The obstacles that make feature selection difficult in GWAS also occur in many other applications of linear regression to big datasets. In fact, the task of finding the optimal subset of features is proven, in general, to be NP-hard \cite{huo2007stepwise}. Therefore, it is usually computationally prohibitive to search over all possible subsets of features and one has to resort to other methods of feature selection. For example, forward (or backward) selection adds (or eliminates) one feature at a time to the regression in a greedy manner \cite{guyon2003introduction}. Alternatively, one may use heuristic methods such as Sure Independence Screening (SIS) \cite{fan2008sure}, which selects features independently based on their correlation with the response, or Minimum Redundancy Maximum Relevance (MRMR) \cite{ding2005minimum}, which penalizes features that are correlated with each other. The most popular approaches to feature selection for linear regression, however, are penalized least-squares methods \cite{hoerl1970ridge,tibshirani1996regression,zou2005regularization,candes2007dantzig} that introduce a penalty function that penalizes large regression coefficients. Common choices for the penalty function include a L2 penalty, called  `Ridge' regression \cite{hoerl1970ridge}, and a L1 penalty, commonly referred to as LASSO regression \cite{tibshirani1996regression}.

Penalized methods for linear regression typically have natural interpretations as Bayesian approaches with appropriately chosen prior distributions.  For example, L2 penalized regression can be derived by maximizing the posterior distribution obtained with a Gaussian prior on the regression coefficients. Similarly, L1 penalized regression can be derived by maximizing the posterior distribution obtained with a Laplace (i.e.\ double-exponential) prior on the regression coefficients. Within a Bayesian framework, relevant features are those with the highest posterior probabilities. However, calculating exact marginal posterior probabilities is generally intractable for high dimensional problems; as a result, the posterior distribution of feature relevance must be explored using Monte Carlo simulations, highlighting the crucial need for new approaches to feature selection \cite{george1993variable,li2010bayesian}.

Inspired by the success of statistical physics approaches to hard problems in computer science \cite{monasson1999determining,mezard2002analytic} and statistics \cite{bialek1996field,nemenman2002occam,balasubramanian1997statistical,malzahn2005statistical,periwal1997reparametrization}, we study high dimensional regression in the ``strongly-regularized regime'' where the prior distribution has a profound influence on the estimators. In the regime where the regression problem is strongly regularized by the prior, we show that the marginal posterior probabilities of feature relevance for L2 penalized regression are well-approximated by the magnetizations of an appropriately chosen Ising model. For this reason, we call our approach the Bayesian Ising Approximation (BIA) of the posterior distribution. Using the BIA, the posterior probabilities can be computed without resorting to Monte Carlo simulation using an efficient mean field approximation that facilitates the analysis of very high dimensional datasets. We envision the BIA as part of a two-stage procedure where the BIA is applied to rapidly screen irrelevant variables, i.e.\ those that have low rank in posterior probability, before applying a more computationally intensive cross validation procedure to infer the regression coefficients for the reduced feature set. Our work is especially well suited to modern feature selection problems where the number of features, $p$, is often larger than the sample size, $n$.

Our approach differs significantly from previous methods for feature selection. Traditionally, penalized regression and related Bayesian approaches have focused on the ``weakly-regularized regime'' where the effect of the prior is assumed to be negligable as the sample size tends to infinity. The underlying intuition for considering the weak-regularization regime is that  as long as the prior (i.e.\ the penalty parameter) is strong enough to regularize the inference problem, a less influential prior distribution should be better suited for feature selection and prediction tasks because it ``allows the data to speak for themselves''. In the machine learning literature, the penalty parameter is usually chosen using cross validation to maximize out-of-sample predictive ability \cite{tibshirani1996regression,zou2005regularization}.  A similar aesthetic is also reflected in the abundant literature on `objective' priors for Bayesian inference \cite{ghosh2011objective}. As expected, these weakly regularizing approaches perform well when the sample size exceeds the number of features ($n \gg p$). However, for high dimensional inference where the number of features can greatly exceed the sample size ($p \gg n$), very strong priors may be required. Our BIA approach exploits the large penalty parameter in this strongly regularized regime to efficiently calculate marginal posterior probabilities using methods from statistical physics.

The paper is organized as follows:  in Section \ref{regress}, we review Bayesian linear regression; in Section \ref{select}, we derive the BIA using a series expansion of the posterior distribution and describe the associated algorithm for variable selection; and in Section \ref{examples} we present (A) analytical results and simulations on the performance of the BIA using features with a constant correlation, (B) we analyze a real dataset for predicting bodyfat percentage from 12 different body measurements, and (C) we analyze a real dataset for predicting a quantitative phenotypic trait from data on the expression of 28,395 genes in soybeans. 

\section{Bayesian Linear Regression}
\label{regress}

In this section, we briefly review the necessary aspects of Bayesian linear regression. This entire section follows standard arguments, the details of which can be found in many textbooks on Bayesian statistics e.g.\ \cite{o2004bayesian}. The goal of linear regression is to infer the set of coefficients $\b_j$ for $j = 1, \ldots, p$ that describe the relationship $y = \bold{x}^T \boldsymbol{\b} + \eta $ from $n$ observations $(y_i, \bold{x}_i)$ for $i = 1, \ldots, n$. Here, $\bold{x}$ is a ($1 \times p)$ vector of features and $\eta \sim \mathcal{N}(0,\s^2)$ is a Gaussian distributed random variable with unknown variance $\s^2$. Without loss of generality, we will assume throughout this paper that the data are standardized with  $ \sum_i y_i = 0$, $\sum_i y_i^2 = n$, $\sum_i (\bold{x}_{i})_j = 0$, and $\sum_i (\bold{x}_{i})_j^2 = n$ so that it is not necessary to include an intercept term in the regression. Penalized least-squares methods estimate the regression coefficients by minimizing a convex objective function in the form of:
\be
U(\boldsymbol{\b}) = \sum_i (y_i - \bold{x}_i^T \boldsymbol{\b})^2 + \l f(\boldsymbol{\b})
\label{eq:errors}
\ee
where $f(\boldsymbol{\b})$ is a function that penalizes large regression coefficients and $\l$ is the strength of the penalty. Common choices for the penalty function include $f(\boldsymbol{\b}) = \sum_j \b_j^2$ for L2 penalized or `Ridge' regression \cite{hoerl1970ridge}, and $f(\boldsymbol{\b}) = \sum_j |\b_j|$ for L1 penalized or LASSO regression \cite{tibshirani1996regression}. The standard least-squares (and maximum likelihood) estimate $\boldsymbol{\hat{\b}} = (X^T X)^{-1} X^T \bold{y}$ is recovered by setting $\l = 0$, where $X$ is the $(n \times p)$ design matrix with rows $\bold{x}_i$. Adding a penalty to the least-squares objective function mitigates instability that results from computing the inverse of the $X^T X$ matrix. In the case of the L1 penalty, many of the regression coefficients end up being shrunk exactly to zero resulting in a type of automatic feature selection \cite{tibshirani1996regression,zou2005regularization,candes2007dantzig}.

Bayesian methods combine the information from the data, described by the likelihood function, with \emph{a priori} knowledge, described by a prior distribution, to construct a posterior distribution that describes one's knowledge about the parameters after observing the data. In the case of linear regression, the likelihood function is a Gaussian:
\be
P(\bold{y} | \boldsymbol{\b}, \s^2) = \left(\frac{1}{\sqrt{2 \pi \s^2}}\right)^n \exp \left( - \frac{(\bold{y} - X^T \boldsymbol{\b})^T (\bold{y} - X^T \boldsymbol{\b})} {2\s^2} \right)
\ee
In this work, we will use standard conjugate prior distributions for $\boldsymbol{\b}$ and $\s^2$ given by $P(\boldsymbol{\b},\s^2 | \bold{s}) = P(\s^2) P(\boldsymbol{\b}|\s^2, \bold{s})$ where:
\begin{align}
P(\s^2) &\propto (\s^2)^{-(a_0 + 1)} \exp(-b_0 / \s^2 ) \\
P(\boldsymbol{\b} | \s^2, \bold{s}) &\propto \prod_j \HF \left[ (1-s_j) \d(\b_j ) + (1+s_j) \sqrt{ \frac{\l}{2 \pi \s^2}} \exp\left(-\frac{\l \b_j^2}{2 \s^2}\right) \right]
\end{align}
These distributions were chosen because they ensure that the posterior distribution can be obtained in closed-form \cite{o2004bayesian}.  Here, we have introduced a vector ($\bold{s}$) of indicator variables so that $\b_j = 0$ if $s_j = -1$ and $\b_j \neq 0$ if $s_j = +1$. We also have to specify a prior for the indicator variables, which we will set to a flat prior $P(\bold{s}) \propto 1$ for simplicity. In principle, $a_0$, $b_0$ and the penalty parameter on the regression coefficients, $\l$, are free parameters that must be specified ahead of time to reflect our prior knowledge. We will discuss these parameters in the next section. 

We have set up the problem so that identifying which features are relevant is equivalent to identifying those features for which $s_j = +1$. Therefore, we need to compute the posterior distribution for $\bold{s}$, which can be determined from Bayes' theorem:
\begin{align}
\log P_{\l}(\bold{s} | \bold{y}) 
&= C + \log \int d\boldsymbol{\b} d\s^2 P(\bold{y} | \boldsymbol{\b}, \s^2) P(\boldsymbol{\b} ,\s^2 |\bold{s}) P(\bold{s}) \nonumber \\
&= C+ \HF \ln | \l I| -\HF \ln | \l I + X _{\bold{s}}^T X_{\bold{s}} | - (a_0 + \frac{n}{2}) \ln (b_0 + \HF E_{\bold{s}}(\l))
\label{eq:posterior}
\end{align}
where $C$ represents a constant and $E_{\bold{s}}(\l)$ is the sum of the squared residual errors. In this expression, $q = \sum_j (1+s_j)/2$, is the number of variables with $s_j = +1$, $I$ is the $(q \times q)$ identity matrix, and $X_{\bold{s}}$ is a $(n \times q)$ restricted design matrix which only contains rows corresponding to features where $s_j = +1$.  The sum of the squared residual errors is given by $E_{\bold{s}}(\l) = \bold{y}^T \bold{y} - \bold{y}^T X_{\bold{s}} \boldsymbol{\bar{\b}}_{\bold{s}}(\l)$, where $\boldsymbol{\bar{\b}}_{\bold{s}}(\l) = (\l I + X_{\bold{s}}^T X_{\bold{s} })^{-1} X_{\bold{s}}^T \bold{y}$ is the Bayesian estimate for the regression coefficients corresponding to those variables for which $s_j = +1$. 

In these expressions, notice that  the Bayesian estimate for the regression coefficients conditioned on $\bold{s}$ (i.e.\ $\boldsymbol{\bar{\b}}_{\bold{s}}(\l)$) is equivalent to an estimate obtained with L2 penalized regression. That is, we have specifically chosen these priors to correspond to a Bayesian formulation of L2 penalized regression. Furthermore, we note that the logarithm of the posterior distribution can be partitioned into an `entropic' term ($\ln | \l I| - \ln | \l I + X _{\bold{s}}^T X_{\bold{s}} |$) measuring the variance of the posterior distribution, and an `energetic' term ($\ln (b_0 + \hf E_{\bold{s}}(\l) ) $) quantifying the fit to the data. 

\section{The Ising Approximation}
\label{select}

\subsection{Strongly Regularized Expansion}

In principle, one can directly use Equation \ref{eq:posterior} to estimate the relevance of each feature using two different approaches. First, we could find the $\bold{s}$ that maximizes the posterior probability distribution. Alternatively, we could compute the marginal probabilities of feature relevance, $P_{\l}(s_j = +1 | \bold{y}) = (1+\<s_j\>) /2$, where  $\<s_j\>$ is the expectation value of $s_j$ with respect to the posterior distribution, and select the features with the largest $P_{\l}(s_j = +1 | \bold{y})$. In the Bayesian setting, these two point estimates result from the use of different utility functions \cite{berger1985statistical}. Here, we will focus on computing the latter, i.e.\ the expected value of $\bold{s}$. The expectation values cannot be evaluated analytically due to the cumbersome restriction of the design matrix to those variables for which $s_j = +1$. Moreover, 
although the computation of the expectation values can be performed using Monte Carlo methods \cite{george1993variable,li2010bayesian}, the numerical calculations often take a long time to converge for high dimensional inference problems. 

Our main result -- which we call the  Bayesian Ising Approximation (BIA) of the posterior distribution for feature selection -- is that a second order series expansion of Equation \ref{eq:posterior} in $\l^{-1}$ corresponds to an Ising model described by:
\begin{align}
\log P_{\l}(\bold{s} | \bold{y}) 
&\simeq C + \frac{n^2}{4 \l}  \left(\sum_{i} h_i(\l) s_i + \HF \sum_{i,j;  i \neq j} J_{ij}(\l) s_i s_j \right) + O\left(\frac{\text{Tr}[ (X_{\bold{s}}^T X_{\bold{s}})^3] }{\l^3}\right)
\label{eq:ising}
\end{align}
where $\text{Tr}[\cdot]$ is the matrix trace operator and the external fields and couplings are defined as:
\begin{align}
h_i(\l) &= r^2(y,x_i) -\frac{1}{n} +\sum_{j} J_{ij}(\l) \label{eq:fields} \\
J_{ij}(\l) &= \frac{n}{\l} \left( \frac{r^2 (x_i,x_j)}{n} -   r(x_i, x_j) r(y,x_i) r(y,x_j) + \HF  r^2(y,x_i) r^2(y,x_j) \right) \label{eq:couplings}
\end{align}
Here, $r(z_1,z_2)$ is the Pearson correlation coefficient between variables $z_1$ and $z_2$.  In writing this expression we have assumed that the hyperparameters $a_0$ and $b_0$ are small enough to neglect, though this assumption is not necessary. A detailed derivation of this result is presented in the Appendix. 

The series expansion converges as long as $\l > \text{Tr}[X_{\bold{s}}^T X_{\bold{s}} ]$ for all $\bold{s}$, which defines the regime that we call `strongly regularized'. Since $X_{\bold{s}}$ is the restricted design matrix for standardized data, we can relate $ \text{Tr}[X_{\bold{s}}^T X_{\bold{s}} ]$ to the covariances between $x_j$'s. In particular, Gershgorin's Circle Theorem \cite{varga2010gervsgorin} implies that the series will converge as long as $\lambda > n(1+ p \tilde{r})$ where $\tilde{r}= \frac{1}{p} \inf_{i} \sum_{j\neq i} |r(X_i, X_j)|$ (see Appendix). For large $p$, we can replace $\tilde{r}$ by the root-mean-squared correlation between features, $r =\sqrt{p^{-1}(p-1)^{-1} \sum_{i \neq j} r^2(X_i, X_j) }$. This defines a natural scale,
\be
\lambda^* = n(1+ p r).
\ee
for the penalty parameter at which the BIA is expected to breakdown. We  expect the BIA to be accurate when $\l \gg \l^*$ and to breakdown when $\l \ll \l^*$.

Because higher order terms in the series can be neglected, the strongly regularized expansion allows us to remove any references to the restricted design matrix, and maps the posterior distribution to the Ising model, which has been studied extensively in the physics literature. To perform feature selection,  we are interested in computing marginal probabilities $P_{\l}(s_j = 1 | \bold{y}) \simeq (1+ m_j(\l) )/2$, where we have defined the magnetizations $m_j(\l) = \< s_j \>$. While there are many techniques for calculating the magnetizations of an Ising model, we focus on the mean field approximation which leads to a self-consistent equation \cite{opper20012}:
\be
m_i(\l) = \tanh \left[  \frac{n^2}{4 \l} \left( h_i(\l) + \HF \sum_{j \neq i} J_{ij}(\l) m_j(\l) \right)  \right]  \label{eq:nmf}
\ee
This mean field approximation provides a computationally efficient tool that approximates Bayesian feature selection for linear regression, requiring only the calculation of the Pearson correlations and solution of Equation \ref{eq:nmf}.

\subsection{Computing the Feature Selection Path}

As with other approaches to penalized regression, our expressions depend on a free parameter ($\l$) that determines the strength of the prior distribution. As it is usually difficult, in practice, to choose a specific value of $\l$ ahead of time it is often helpful to compute the feature selection path; i.e.\ to compute $m_j(\l)$ over a wide range of $\l$'s. Indeed, computing the variable selection path is a common practice when applying other feature selection techniques such as LASSO regression. To obtain the mean field variable selection path as a function of $\e = 1/\l$, we notice that $\lim_{\e \to 0} m_j(\e) = 0$ and so define the recursive formula:
\be
m_i \left(\e +\d\e\right) \approx \tanh \left[  \frac{(\e + \d\e) n^2}{4} \left( h_i\left(\e + \d\e\right) + \HF \sum_{j \neq i} J_{ij}\left(\e + \d\e\right) m_j\left(\e\right) \right)  \right]  \label{eq:recursive}
\ee
with a small step size $\d \e \ll 1/\l^{*} = n^{-1}( 1+ pr)^{-1}$. We have set $\d \e = 0.05 / \l^{*}$ in all of the examples presented below. 

\subsection{Remarks}

The BIA provides a computationally efficient framework to calculate posterior probabilities of feature relevance as a function of $\l$ without Monte Carlo simulations.  The local fields and couplings of the BIA (Eqs.\ \ref{eq:fields}, \ref{eq:couplings}) are simple functions of Pearson correlation coefficients. The most challenging computational aspect of feature selection with the BIA is the large amount of memory required for storing the $(p \times p)$ coupling matrix for very high dimensional problems.  One potential route for decreasing the memory requirement is to use adaptive thresholding estimators for the correlations to obtain a sparse coupling matrix, though we do not explore this idea further in this work because memory requirements did not cause a problem for our examples even when considering datasets with $p \sim 30,000$ features. 

To first order in $\e= \l^{-1}$, the posterior distribution corresponds to an Ising model with fields and couplings given by $h_i = r^2(y,x_i) - 1/n$ and $J_{ij} = 0$. That is, the spin variables representing feature relevance are independent, and the probability that a feature is relevant is only a function of its squared correlation with the response. Specifically, $m_j(\l) \geq 0$ if $ |r(y,x_j)| > 1/\sqrt{n}$ and $m_j(\l) \leq 0$ if $ |r(y,x_j)| < 1/\sqrt{n}$. Therefore, the BIA demonstrates that methods that rank features by their squared Pearson correlation with the response, such as Sure Independence Screening \cite{fan2008sure}, are actually performing a first order approximation to Bayesian feature selection in the strongly regularized limit. 

The couplings between the spin variables representing feature relevance enter into the BIA with the second order term in $\epsilon= \l^{-1}$. A positive (or `ferrogmagnetic') coupling between spins $i$ and $j$ favors models that include both features $i$ and $j$, whereas a negative (or `antiferromagnetic') coupling favors models that include one feature or the other, but not both. In general, the coupling terms are antiferromagnetic for highly correlated variables, which minimizes the redundancy of the feature set. 

\section{Examples}
\label{examples}

We have chosen three examples to illustrate different characteristics of the BIA for Bayesian feature selection. (A) First, we consider regression problems with $p$ features that have a constant correlation $r$. We present some simple analytic expressions in the large $p$ limit that illustrate how different aspects of the problem affect feature selection performance, and study some simulated data. (B) Next, we analyze a dataset on the prediction of bodyfat percentage from various body measurements. The number of features ($p = 12$) is small enough that we can compute the exact posterior probabilties and, therefore, directly assess the accuracy of the BIA for these data. (C) Finally, we demonstrate the applicability of the BIA for feature selection on high dimensional regression problems by examining a dataset relating the expression of $p = 28395$ genes to the susceptibility of soybean plants to a pathogen. 

\subsection{Features with a Constant Correlation}

Intuitively, one may expect that correlations between features are detrimental to feature selection. Indeed, previous observations on feature selection with LASSO have demonstrated the negative impact of inter-feature correlations on variable selection performance \cite{tibshirani1996regression,zou2005regularization}. Given these observations, we use this section to analyze a simple model of BIA feature selection that allows us to examine many of the characteristics that influence feature selection performance. Specifically, we consider a simple, analytically tractable, model in which we are given $p$ features that are correlated with each other with a constant Pearson correlation coefficient, $r$. The response, $\tilde{y}$, is a linear function of the first $\tilde{p} \leq p$ variables, which have equal true regression coefficients $\b_j = \b$ for $j \leq \tilde{p}$. That is, $\tilde{y} = \b \sum_{j=1}^{j=\tilde{p}} x_j + \tilde{\eta}$ where $\tilde{\eta} \sim \mathcal{N}(0,\tilde{\s}^2)$ is a Gaussian noise. We are interested in studying the behavior of this model when the number of features is large ($p \gg 1$). To simplify analytic expressions, it is helpful to define the number of samples as $n = \theta p$, and the number of relevant features as $\tilde{p} = \phi p$. Furthermore, we assume that the correlation between features scales as $r= \alpha p^{-1}$ so that the correlation between $y$ and $x_j$ stays constant in the large $p$ limit.

\begin{figure}[t]
\begin{center}
  \includegraphics[angle=0,width=0.75\textwidth]{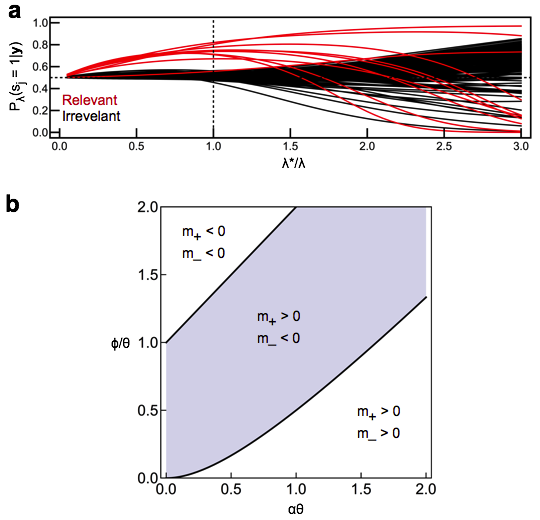}
  \caption{\label{fig:fig1} Performance of BIA feature selection.  a) An example variable selection path as a function of decreasing regularization. The relevant variables are red, and the irrelevant variables are black. The dashed vertical line is at $ \l = \l^* = n(1+rp)$, which is the estimated breakdown point of the approximation. Simulations were performed with $p=200$, $n = 100$, $\tilde{p} = 10$, $r = 2/p$, and $\w^2 = 1$. b) A phase diagram illustrating the regions of parameter space where $m_{(-)} < 0 < m_{(+)}$ computed with $\l = \theta p^2$.   }
\end{center}
\end{figure}

Figure \ref{fig:fig1}a presents an example feature selection path computed using the BIA for a simulation of this model. This variable selection path was generated for data simulated from a linear model using with $p=200$ features with a constant correlation $r = 2/p$, $n = 100$, $\tilde{p} = 10$, and $\w^2 = \tilde{\s}^2 / \b^2 = 1$.  Figure \ref{fig:fig1}a demonstrates that all but one of the relevant features (red) have higher posterior probabilities than the irrelevant features (black) as long as $\l > \l^*$. In fact, there is a clear gap in posterior probability separating the relevant and irrelevant features, and the correct features can be easily selected by visible inspection of the feature selection path in Figure \ref{fig:fig1}a. The BIA breaks down beyond the threshold of the penalty parameter and the feature selection performance of the BIA deteriorates, as demonstrated by the mixing of the probabilities for the relevant (red lines) and irrelevant (black lines) features in Figure \ref{fig:fig1}a.

Of course, we expect that the performance of the BIA for feature selection will vary depending on the characteristics of the problem. The model with equally correlated features provides a simple scenario to study which characteristics affect feature selection performance, because the correlations between the features and the standardized response $y = \tilde{y} / \sqrt{\text{VAR}(\tilde{y})}$ can be easily computed analytically for the large sample size limit where we can neglect sample-to-sample fluctuations. 

The spins characterizing the feature selection problem can be divided into two groups: relevant features with $j \leq \tilde{p}$ and magnetization $m_{(+)}$, and irrelevant features with $j > \tilde{p}$ and magnetization $m_{(-)}$. Note that an algorithm that performs perfect variable selection will have $m_{(+)} = + 1$ and $m_{(-)} = -1$. The Pearson correlation coefficient of a relevant feature ($j \leq \tilde{p}$) with the standardized response $y = \tilde{y} / \sqrt{\text{VAR}(\tilde{y})}$ is given by:
\be
r(y,x_{j=1\ldots \tilde{p}}) \equiv r_{(+)} = \frac{ 1 + r (\tilde{p}-1) } {\sqrt{ \w^2 + \tilde{p} (r \tilde{p} + 1 - r) } }  \nonumber 
\ee
where $\w^2 = \tilde{\s}^2 / \b^2 \sim O(1)$ is an inverse signal-to-noise ratio. Similarly, the Pearson correlation coefficient of an irrelevant variable ($j > \tilde{p}$) with the standardized response is:
\be
r(y,x_{j=\tilde{p}+1, \ldots, p}) \equiv r_{(-)} = \frac{r \tilde{p}} {\sqrt{ \w^2 + \tilde{p} (r \tilde{p} + 1 - r) } }  \nonumber 
\ee
If we choose $\l = \theta p^2$ to ensure that the problem is always in the strongly regularized regime, the magnetizations can be computed explicity to order $1/p$ giving:
\begin{align}
m_{(+)} &\approx \frac{\theta - \phi(1 - \a \theta) }{4 \phi} \frac{1}{p} + O(\frac{1}{p^2}) \nonumber \\
m_{(-)} &\approx -\frac{1+\a \phi - \a^2 \phi \theta}{4(1+\a \phi) } \frac{1}{p} + O(\frac{1}{p^2}) \nonumber
\end{align}

In general, we say that feature selection performance is good, on average, as long as $m_{(-)} < 0 < m_{(+)}$, because revelant features have $P(s_j=+1|\bold{y}) > 1/2$ and irrelevant features have $P(s_j=+1|\bold{y}) < 1/2$. Figure \ref{fig:fig1}b shows that the average feature selection performance is good in this sense within a large volume of the phase space. Specifically,  $m_{(-)} < 0 < m_{(+)}$ when:
\be
\frac{1}{1+\a \phi} < \frac{\theta}{\phi} < \frac{1+\a \phi}{(\phi \a)^2} \nonumber
\ee
However, $m_{(-)}< m_{(+)}$ even if the stronger statement $m_{(-)} < 0 < m_{(+)}$ is not satisfied. As a result, there is always a gap between the posterior probabilities of the relevant and irrelevant features. 

Feature selection is most difficult if the features are correlated and if the number of relevant features is large compared to the sample size. Moreover, note that our choice of $\l = \theta p^2$ leads to $|m_{(\pm)} | \ll 1$ in the large $p$ limit, indicating a high degree of uncertainty even in the regime in which the BIA is accurate and in which the signs of the magnetizations are correct. Our choice of $\l = \theta p^2$ provides much stronger regularization than the estimated breakdown point of $\l^{*} = \theta (1+\a)p$. As a result, the absolute magnitudes of the magnetizations (i.e.\ $|m_{(\pm)}|$) are small, even compared to other values of $\l$ for which the BIA still holds. 

\subsection{Bodyfat Percentage}

\begin{figure}[t]
\begin{center}
  \includegraphics[angle=0,width=0.75\textwidth]{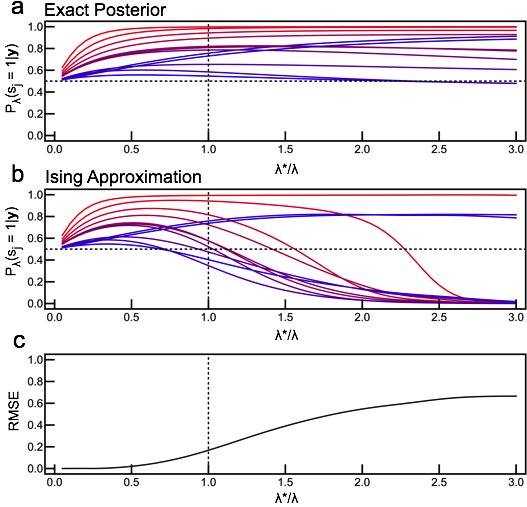}
  \caption{\label{fig:fig2} Comparison of exact Bayesian marginal probabilities to the BIA for the bodyfat data. a) Exact Bayesian marginal probabilities for decreasing regularization. b) BIA approximations of the marginal probabilities for decreasing regularization. c) Root Mean Squared Error (RMSE) between the exact and BIA probabilities as a function of decreasing regularization. The dashed vertical line is at $ \l = \l^* = n(1+rp)$, which is the estimated breakdown point of the approximation. The variables have been color coded (blue to red) by increasing squared Pearson correlation coefficient with the response (bodyfat percentage).  }
\end{center}
\end{figure}

Bodyfat percentage is an important indicator of health, but obtaining accurate estimates of bodyfat percentage is challenging. For example, underwater weighing is one of the most accurate methods for measuring bodyfat percentage but it requires special equipment, e.g.\ a pool. Here, we analyze a well-known dataset obtained from StatLib (http://lib.stat.cmu.edu/datasets/) on the relationship between bodyfat percentage and various body measurements from $n = 252$ men \cite{penrose1985generalized}. The $p = 12$ features included in our regression are: age and body mass index ($\text{height} / \text{mass}^2$), as well as circumference measurements of the neck, chest, waist, hip, thigh, knee, ankle, upper arm, forearm and wrist. All of the data were standardized to have mean zero and variance one. Therefore, there are $2^{12} = 4096$ potential combinations of features.

For our purposes, the most interesting part about the bodyfat dataset is that the number of features is small enough to compute the posterior probabilities exactly using Equation \ref{eq:posterior} by enumerating all of the $4096$ feature combinations. The exact posterior probabilities as a function of $\l^{-1}$ are shown in Figure \ref{fig:fig2}a. In the figure, we have color coded variables from blue to red in terms of increasing squared Pearson correlation coefficients with bodyfat percentage; (blue) ankle, body mass index, age, wrist, forearm, neck, upper arm, knee, thigh, hip, chest, waist (red). The posterior probabilities computed from the BIA (Equation \ref{eq:recursive}) are shown in Figure \ref{fig:fig2}b.

Comparing Figures \ref{fig:fig2}a-b demonstrates that the posterior probabilities computed from the BIA are very accurate for $\l \gg  \l^*$, with $\l^*=n(1+pr)$ and $r$  the root-mean-squared correlation between features. However, the approximation breaks down for $\l \ll \l^*$ as expected. Figure \ref{fig:fig2}c provides another representation of the breakdown of the BIA upon approaching the breakdown point of the penalty ($\l^*$). The Root Mean Squared Error given by $\text{RMSE}(\l) =\sqrt{p^{-1} \sum_j (P_{\l}^{\text{exact}}(s_j = 1 | \bold{y} )- P_{\l}^{\text{BIA}}(s_j = 1 | \bold{y}))^2}$ is sigmoidal, with an inflection point close to $\l^*$. 

In the strongly regularized regime with $\l \gg \l^*$, the exact Bayesian probabilties and those computed using the BIA both rank waist and chest circumference as the most relevant features. Below the breakdown point of the penalty parameter, however, the BIA suggests solutions that are too sparse. That is, it underestimates many of the posterior probabilities describing whether or not the features are relevant. Far below the breakdown point of the penalty parameter (beyond the range of the graph in Figure \ref{fig:fig2}), the BIA ranks age and body mass index as the most relevant variables even though these have some of the smallest correlations with the response. Age and body mass index also become increasingly important for small $\l$'s in the exact calculation; though, they are never ranked as the most relevant variables. The change in the rankings of the features as a function of $\l$ highlights the importance of the coupling terms ($J_{ij}(\l)$) that punish correlated features. 

\subsection{Gene Expression}

In 2010, the Dialogue for Reverse Engineering Assessments and Methods (DREAM) \cite{prill2010towards} initiative issued a challenge to predict the response of soybean plants to a pathogen from data on gene expression \cite{zhou2009infection}. The training data consist of a response of $n = 200$ different soybean plants to a pathogen along with the expressions of $p = 28395$ genes. The team (Loh et al.\ \cite{loh2011phenotype}) that achieved the highest rank correlation on a blind test set of 30 other soybean plants trained their model using elastic net regression to predict the ranks of the responses in the training set. The ranks were used rather than the actual values of the responses to mitigate the effects of outliers, and the value of the penalty parameter was chosen using cross validation. Loh et al.\ found that their cross validation procedure for elastic net regression favored sparse models with only a few features, and they  highlighted 12 of these features that were frequently chosen by their procedure \cite{loh2011phenotype}.    

Clearly, this DREAM-5 soybean dataset presents a severely underdetermined problem, with the number of features exceeding the sample size by two orders of magnitude. Therefore, it is unsurprising, perhaps, that even the best teams achieved only modest performance on the test data \cite{loh2011phenotype}. Nevertheless, the soybean gene expression dataset presents a good benchmark to compare Bayesian feature selection with the BIA to feature selection using cross validated penalized regression for a very high dimensional inference problem. 

\begin{figure}[t]
\begin{center}
  \includegraphics[angle=0,width=0.75\textwidth]{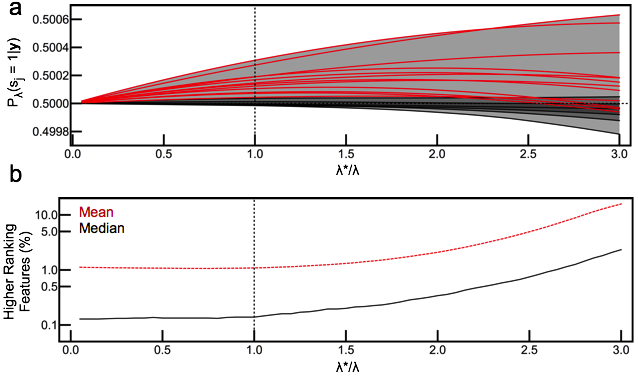}
  \caption{\label{fig:fig3} Feature selection path for the gene expression data. The problem is severely under-determined, involving the prediction of a quantitative phenotype from the expressions of $p=28395$ genes given a sample size of $n=200$ and, therefore, the posterior probabilities remain close to $P_{\l}(s_j = 1 | \bold{y}) = 1/2$. a) Features selected in a previous study (red lines) by cross validation with the elastic net have high ranking posterior probabilities. Gray scale represents the outer 10\% quantiles (light gray), the outer 10\% - 25\% quantiles (gray), and the middle 50\% quantiles (dark gray). b) The median (solid black line) and mean (dashed red line) percentage of features with higher posterior probabilities than those identified by Loh et al. The vertical axis is a logarithmic scale. The dashed vertical line is at $ \l = \l^* = n(1+rp)$, which is the estimated breakdown point of the approximation. }
\end{center}
\end{figure}

We used the BIA to compute the posterior probabilities for all $p = 28395$ features as a function of $\l^{-1}$.  Following the lead from the team that won the DREAM-5 challenge, we chose our $y$ variable as the ranks of the responses of the soybean plants to the pathogen rather than the actual values. As before, all of the data were standarized to have mean zero and variance one. It is not particularly helpful to plot the posterior probabilities of all $28395$ features. Therefore, Figure \ref{fig:fig3}a compares the posterior probabilities of the 12 features highlighted by Loh et al.\ (red lines) to the distribution of posterior probabilities for all of the features (gray area). Here, the distribution of posterior probabilities is represented by quantiles; the gray scale represents the outer 10\% quantiles (light gray), the outer 10\% - 25\% quantiles (gray), and the middle 50\% quantiles (dark gray). Visual inspection of Figure \ref{fig:fig3}a suggests that the 12 features identified by Loh et al.\ have some of the highest posterior probabilities among all $28395$ features. Similarly, Figure \ref{fig:fig3}b shows that only a small percentage of features have higher posterior probabilities than those identified by Loh et al, demonstrating that there is generally pretty good agreement between features that are predictive (i.e.\ those that perform well in cross validation) and those with high posterior probabilities computed with the BIA.

Although our analyses of the soybean gene expression data identifies similar features as cross validated elastic net regression, the posterior probabilities all fall in the range $P_{\l}(s_j | \bold{y}) = 1/2 \pm 0.001$. The small range of posterior probabilities around the value representing random chance ($P_{\l}(s_j | \bold{y}) = 1/2$) that we identify is consistent with the highly variable out-of-sample performance discussed by Loh et al. The fact that there is no strong evidence favoring the selection of any of the features is not really surprising considering the vastly underdetermined nature of the problem ($n = 200$ and $p = 28395$). Moreover, the gene expression features have a root mean square correlation of $r \approx 0.29$. As a result, the critical value of the penalty parameter is $\l^* = n(1+pr) \approx 1.65 \times 10^6$, which is huge compared to what the breakdown point of $\l^* = 200$ would be if we were to assume that $r = 0$. Given that smaller $\l$'s generally lead to magnetizations that are larger in absolute value, it is clear that ignoring the correlations between genes vastly inflates estimates of certainty in gene relevance. This highlights the importance of strong regularization procedures that specifically account for correlation between genes in high dimensional genomic studies.

\section{Discussion} 

To summarize, we have shown that Bayesian feature selection for L2 penalized regression, in the strongly regularized regime, corresponds to an Ising model, which we call the Ising Approximation (BIA). Mapping the posterior distribution to an Ising model that has simple expressions for the local fields and couplings using a controlled approximation opens the door to analytical studies of Bayesian feature selection using the vast number of techniques developed in physics for studying the Ising model. In fact,  our analyses can be generalized to study Bayesian feature selection for many statistical techniques other than linear regression, as well as other prior distributions \cite{FisherBayesian2}. 
From a practical standpoint, the BIA provides an algorithm to efficiently compute Bayesian feature selection paths for L2 penalized regression. Using our approach, it is possible to compute posterior probabilities of feature relevance for very high dimensional datasets such as those typically found in genomic studies. 

Unlike most previous work of feature selection, the BIA is ideally suited for large genomic datasets where the number of features can be much greater than the sample size, $p \gg n$. The underlying reason for this is that we work in strongly-regularized regime where the prior always has a large influence on the posterior probabilities. This is in contrast to previous works on penalized regression and related Bayesian approaches that have focused on the ``weakly-regularized regime'' where the effect of the prior is assumed to be small. Moreover, we have identified a sharp threshold for the regularization parameter $\l^* = n(1+pr)$  where the BIA is expected to break down. This threshold depends on the sample size, $n$, number of features, $p$, and root-mean-squared correlation between features, $r$. The threshold at which the BIA breaks down occurs precisely at the transition from the strongly-regularized to the weakly-regularized regimes where the prior and the likelihood have a comparable influence on the posterior distribution.

Our work also highlights the importance of  accounting for correlations between features when assessing statistical significance in large data sets. In general, we have found that when the number of features is large, even small correlations can cause a huge reduction in the posterior probabilities of features. For example, our analysis of a dataset including the expression of 28,395 genes in soybeans demonstrates that the resulting posterior probabilities of gene relevance may be very close to value representing random chance $P_{\l}(s_j | \bold{y}) = 1/2$ when $p \gg n$ and the genes are moderately correlated, e.g.\ $r \sim 0.29$. This is likely to have important implications for assessing the results of GWAS studies where such correlations are often ignored.

Another implication of the small marginal posterior probabilities resulting from correlations among potential features is that  it is probably not reasonable to choose a posterior probability threshold for judging significance on very high dimensional problems. Instead, the BIA can be used as part of a two-stage procedure in the same manner as Sure Independence Screening \cite{fan2008sure}, where the BIA is applied to rapidly screen irrelevant variables, i.e.\ those that have low rank in posterior probability, before applying a more computationally intensive cross validation procedure to infer the regression coefficients. The computational efficiency of the BIA and the existence of a natural threshold  for the penalty parameter where the BIA works  make this procedure ideally suited for such two stage procedures.

\section{Acknowledgements} 

\section{Appendix}

\setcounter{equation}{0}
\setcounter{figure}{0}
\makeatletter 
\renewcommand{\theequation}{S\@arabic\c@equation} 
\makeatletter 
\renewcommand{\thefigure}{S\@arabic\c@figure} 
\makeatletter 
\renewcommand{\theequation}{S\@arabic\c@equation} 

\subsection{Bayesian Linear Regression}

In this section, we briefly review the necessary aspects of Bayesian linear regression. This entire section follows standard arguments, the details of which can be found in many textbooks on Bayesian statistics e.g.\ \cite{o2004bayesian}, and also appears in the main text; we repeat it here so that the Appendix is self contained. The goal of linear regression is to infer the set of coefficients $\b_j$ for $j = 1, \ldots, p$ that describe the relationship $y = \bold{x}^T \boldsymbol{\b} + \eta $ from $n$ observations $(y_i, \bold{x}_i)$ for $i = 1, \ldots, n$. Here, $\bold{x}$ is a ($1 \times p)$ vector of features and $\eta \sim \mathcal{N}(0,\s^2)$ is a Gaussian distributed random variable with unknown variance $\s^2$. Without loss of generality, we will assume throughout this paper that the data are standardized with  $ \sum_i y_i = 0$, $\sum_i y_i^2 = n$, $\sum_i (\bold{x}_{i})_j = 0$, and $\sum_i (\bold{x}_{i})_j^2 = n$ so that it is not necessary to include an intercept term in the regression. Penalized least-squares methods estimate the regression coefficients by minimizing a convex objective function in the form of:
\be
U(\boldsymbol{\b}) = \sum_i (y_i - \bold{x}_i^T \boldsymbol{\b})^2 + \l f(\boldsymbol{\b})
\ee
where $f(\boldsymbol{\b})$ is a function that penalizes large regression coefficients and $\l$ is the strength of the penalty. Common choices for the penalty function include $f(\boldsymbol{\b}) = \sum_j \b_j^2$ for L2 penalized or `Ridge' regression \cite{hoerl1970ridge}, and $f(\boldsymbol{\b}) = \sum_j |\b_j|$ for L1 penalized or LASSO regression \cite{tibshirani1996regression}. The standard least-squares (and maximum likelihood) estimate $\boldsymbol{\hat{\b}} = (X^T X)^{-1} X^T \bold{y}$ is recovered by setting $\l = 0$, where $X$ is the $(n \times p)$ design matrix with rows $\bold{x}_i$. Adding a penalty to the least-squares objective function mitigates instability that results from computing the inverse of the $X^T X$ matrix. In the case of the L1 penalty, many of the regression coefficients end up being shrunk exactly to zero resulting in a type of automatic feature selection \cite{tibshirani1996regression,zou2005regularization,candes2007dantzig}.

Bayesian methods combine the information from the data, described by the likelihood function, with \emph{a priori} knowledge, described by a prior distribution, to construct a posterior distribution that describes one's knowledge about the parameters after observing the data. In the case of linear regression, the likelihood function is a Gaussian:
\be
P(\bold{y} | \boldsymbol{\b}, \s^2) = \left(\frac{1}{\sqrt{2 \pi \s^2}}\right)^n \exp \left( - \frac{(\bold{y} - X^T \boldsymbol{\b})^T (\bold{y} - X^T \boldsymbol{\b})} {2\s^2} \right)
\ee
In this work, we will use standard conjugate prior distributions for $\boldsymbol{\b}$ and $\s^2$ given by $P(\boldsymbol{\b},\s^2 | \bold{s}) = P(\s^2) P(\boldsymbol{\b}|\s^2, \bold{s})$ where:
\begin{align}
P(\s^2) &\propto (\s^2)^{-(a_0 + 1)} \exp(-b_0 / \s^2 ) \\
P(\boldsymbol{\b} | \s^2, \bold{s}) &\propto \prod_j \HF \left[ (1-s_j) \d(\b_j ) + (1+s_j) \sqrt{ \frac{\l}{2 \pi \s^2}} \exp\left(-\frac{\l \b_j^2}{2 \s^2}\right) \right]
\end{align}
These prior distributions were chosen so that the posterior distribution has a simple closed-form expression. Here, we have introduced a vector ($\bold{s}$) of indicator variables so that $\b_j = 0$ if $s_j = -1$ and $\b_j \neq 0$ if $s_j = +1$. We also have to specify a prior for the indicator variables, which we will set to a flat prior $P(\bold{s}) \propto 1$ for simplicity. In principle, $a_0$, $b_0$ and the penalty parameter, $\l$, are free parameters that must be specified ahead of time and reflect our prior knowledge. We will discuss these parameters more in the next section.

We have set up the problem so that identifying which features are relevant is equivalent to identifying those features for which $s_j = +1$. Therefore, we need to compute the posterior distribution for $\bold{s}$, which can be determined from Bayes' theorem:
\begin{align}
\log P_{\l}(\bold{s} | \bold{y}) 
&= C + \log \int d\boldsymbol{\b} d\s^2 P(\bold{y} | \boldsymbol{\b}, \s^2) P(\boldsymbol{\b} ,\s^2 |\bold{s}) P(\bold{s}) \nonumber \\
&= C+ \HF \ln | \l I| -\HF \ln | \l I + X _{\bold{s}}^T X_{\bold{s}} | - (a_0 + \frac{n}{2}) \ln (b_0 + \HF E_{\bold{s}}(\l))  \nonumber \\
&\equiv \mathcal{L}(\bold{s} | y) 
\end{align}
where $C$ represents a constant and $E_{\bold{s}}(\l)$ is the sum of the squared residual errors. In this expression, $q = \sum_j (1+s_j)/2$, is the number of variables with $s_j = +1$, $I$ is the $(q \times q)$ identity matrix, and $X_{\bold{s}}$ is a $(n \times q)$ restricted design matrix which only contains rows corresponding to features where $s_j = +1$.  The sum of the squared residual errors is given by $E_{\bold{s}}(\l) = \bold{y}^T \bold{y} - \bold{y}^T X_{\bold{s}} \boldsymbol{\bar{\b}}_{\bold{s}}(\l)$, where $\boldsymbol{\bar{\b}}_{\bold{s}}(\l) = (\l I + X_{\bold{s}}^T X_{\bold{s} })^{-1} X_{\bold{s}}^T \bold{y}$ is the Bayesian estimate for the regression coefficients corresponding to those variables for which $s_j = +1$. 

\subsection{Strongly Regularized Expansion}

Now, we will perturbatively study the model selection posterior distribution about the limit where $\l$ is large. It is helpful to rewrite the expressions in terms of $\e = 1/\l$ which will be the small parameter that we will use in the expansion. The log-posterior is:
\begin{align}
\mathcal{L}(\bold{s} | y) 
&= \text{constant} + \HF \ln |I| - \HF \ln | I +  \e X_{\bold{s}}^T X_{\bold{s}} | - (a_0 + \frac{n}{2}) \ln (b_0 + \hf y^Ty - \hf y^T X_{\bold{s}} \boldsymbol{\bar{\b}}_{\bold{s}}(\e) ) \nonumber
\end{align}
where $\boldsymbol{\bar{\b}}_{\bold{s}}(\e) = \e (I + \e X_{\bold{s}}^T X_{\bold{s} })^{-1} X_{\bold{s}}^T \bold{y}$. We will expand $\mathcal{L}(\bold{s}|y)$ in powers of $\e$ to second order. For now, we will assume that higher order terms can be neglected, and we will ask later on when this assumption breaks down. We need:
\be
\ln | I + \e X_{\bold{s}}^T X_{\bold{s}} | - \ln | I | = \e \text{Tr}[X_{\bold{s}}^T X_{\bold{s}}] - \HF \e^2 \text{Tr}[ (X_{\bold{s}}^T X_{\bold{s}})^2] + O( \e^3 ) \nonumber 
\ee
and
\begin{align}
&\ln (b_0 + \frac{1}{2}( y^Ty - y^T X_{\bold{s}} \boldsymbol{\bar{\b}}_{\bold{s}}(\e) ) + \ln 2 - \ln (2b_0 + n)\nonumber \\
&= \e \left( \frac{ y^T X_{\bold{s}} \del_{\e} \boldsymbol{\bar{\b}}_{\bold{s}}(0) } {2 b_0 + n} \right) - \frac{\e^2}{2} \left( \frac{ y^T X_{\bold{s}} \del_{\e}^2 \boldsymbol{\bar{\b}}_{\bold{s}}(0) }{2 b_0 + n} + \left( \frac{y^T X_{\bold{s}} \del_{\e} \boldsymbol{\bar{\b}}_{\bold{s}}(0) }{2 b_0 +  n } \right)^2 \right) + O(\e^3 )  \nonumber
\end{align}
Now, we can calculate:
\begin{align}
\del_{\e} \boldsymbol{\bar{\b}}_{\bold{s}}(0)
&= \del_{\e}  \left[\e ( I + \e X_{\bold{s}}^T X_{\bold{s}})^{-1} X_{\bold{s}}^T y \right]_{\e=0} \nonumber \\
&= \left[( I + \e X_{\bold{s}}^T X_{\bold{s}})^{-1} X_{\bold{s}}^T y - \e ( I + \e X_{\bold{s}}^T X_{\bold{s}})^{-1} X_{\bold{s}}^T X_{\bold{s}} ( I + \e X_{\bold{s}}^T X_{\bold{s}})^{-1} X_{\bold{s}}^T y \right]_{\e=0} \nonumber \\
&= X_{\bold{s}}^T y \nonumber
\end{align}
and
\begin{align}
\del_{\e}^2 \boldsymbol{\bar{\b}}_{\bold{s}}(0)
&= \del_{\e}^2  \left[\e ( I + \e X_{\bold{s}}^T X_{\bold{s}})^{-1} X_{\bold{s}}^T y \right]_{\e=0} \nonumber \\
&=  \left[\del_{\e}  ( I + \e X_{\bold{s}}^T X_{\bold{s}})^{-1} X_{\bold{s}}^T y - \del_{\e}  \e ( I + \e X_{\bold{s}}^T X_{\bold{s}})^{-1} X_{\bold{s}}^T X_{\bold{s}} ( I + \e X_{\bold{s}}^T X_{\bold{s}})^{-1} X_{\bold{s}}^T y \right]_{\e=0} \nonumber \\
&=  -2 \left[ ( I + \e X_{\bold{s}}^T X_{\bold{s}})^{-1} X_{\bold{s}}^T X_{\bold{s}} ( I + \e X_{\bold{s}} X_{\bold{s}})^{-1} X_{\bold{s}}^T y\right]_{\e=0} \nonumber \\
&-\left[ \e \del_{\e} ( I + \e X_{\bold{s}}^T X_{\bold{s}})^{-1} X_{\bold{s}}^T X_{\bold{s}} ( I + \e X_{\bold{s}}^T X_{\bold{s}})^{-1} X_{\bold{s}}^T y \right]_{\e=0} \nonumber \\
& = -2 X_{\bold{s}}^T X_{\bold{s}} X_{\bold{s}}^T y \nonumber
\end{align}
Therefore, (up to a constant term):
\begin{align}
&\ln (b_0 + \frac{1}{2}( y^T y - y^T X_{\bold{s}} \boldsymbol{\bar{\b}}_{\bold{s}}(\e) ) \nonumber \\
&=  - \e \left( \frac{ y^T X_{\bold{s}} X_{\bold{s}}^T y } {2 b_0 + n} \right) + \e^2\left( \frac{y^T X_{\bold{s}} X_{\bold{s}}^T X_{\bold{s}} X_{\bold{s}}^T y }{2 b_0 + n} - \HF \left( \frac{y^T X_{\bold{s}} X_{\bold{s}}^T y }{2 b_0 +  n} \right)^2 \right) + O(\e^3) \nonumber 
\end{align}
Putting things together the log-posterior is (up to a constant term):
\begin{align}
\mathcal{L}(\bold{s} | y) 
&= \frac{\e}{2} \left( \left(\frac{2 a_0 + n}{ 2 b_0 + n} \right) y^T X_{\bold{s}} X_{\bold{s}}^T y - \text{Tr}[X_{\bold{s}}^T X_{\bold{s}}] \right)  \nonumber \\
&+ \frac{\e^2}{2} \left( \text{Tr}[(X_{\bold{s}}^T X_{\bold{s}})^2] - \left(\frac{2 a_0 + n}{2 b_0 + n} \right) \left( y^T X_{\bold{s}} X_{\bold{s}}^T X_{\bold{s}} X_{\bold{s}}^T y - \HF \frac{(y^T X_{\bold{s}} X_{\bold{s}}^T y)^2 }{2 b_0 +  n} \right) \right) + O(\e^3) \nonumber
\end{align}
To simplify things a little, we will assume that $a_0$ and $b_0$ can be neglected, which gives:
\begin{align}
\mathcal{L}(\bold{s} | y) 
&= \frac{\e}{2} \left( y^T X_{\bold{s}} X_{\bold{s}}^T y - \text{Tr}[X_{\bold{s}}^T X_{\bold{s}}] \right)  \nonumber \\
&+ \frac{\e^2}{2} \left( \text{Tr}[(X_{\bold{s}}^T X_{\bold{s}})^2] -  \left( y^T X_{\bold{s}} X_{\bold{s}}^T X_{\bold{s}} X_{\bold{s}}^T y - \frac{1}{2n} (y^T X_{\bold{s}} X_{\bold{s}}^T y)^2 \right) \right) + O(\e^3)
\label{eq:expanded-posterior}
\end{align}

\subsection{Breakdown of the Approximation}

In a truly Bayesian setting, the parameters $a_0$, $b_0$ and $\l$ are chosen ahead of time to reflect the prior knowledge of the statistician. By contrast, L2 penalized regression is also commonly used in a frequentist setting with $\l$ chosen by cross-validation. In any case, the inclusion of the penalty parameter helps to regularize the inverse of $X_{\bold{s}}^T X_{\bold{s}}$, which is often of low rank. Indeed, in the high-dimensional setting with $p > n$ the $(p \times p)$ matrix $X_{\bold{s}}^T X_{\bold{s}}$ has a maximum rank of $n$ and is, therefore, never invertible. Note, however, that we can always compute the inverse of $\Lambda = (\l I + X_{\bold{s}}^T X_{\bold{s}})$ for any $\l > 0$ because the combination of a postive definite matrix with a postive semi-definite matrix is postive definite. The convergence of the series expansion for $\Lambda$ is, by and large, the factor determining the convergence of the series expansion for the log-posterior. Let's expand the inverse of $\Lambda$ about $\l = \infty$ as
\be
\Lambda^{-1} = \l^{-1}( I + \l^{-1} X_{\bold{s}}^T X_{\bold{s}})^{-1} = \l^{-1} \sum_{k = 0}^{\infty} (-1)^k \l^{-k} (X_{\bold{s}}^T X_{\bold{s}})^k \nonumber
\ee
The geometric series converges as long as $\l > \text{Tr}[ X_{\bold{s}}^T X_{\bold{s}} ]$, and truncating the series after the $k^{th}$ order term leads to an error of order $O( \text{Tr}[ (X_{\bold{s}}^T X_{\bold{s}} / \l )^{(k+1)} ] )$. Thus, to ensure that the series converges for all $\bold{s}$ we need $\l > \text{Tr}[ X^T X ]$, where $X$ is the design matrix for all $p$ features. 

For large $k$, we know that  $Tr[(X_{\bold{s}}^T X_{\bold{s}} )^{(k+1)}]$ is dominated by the largest eigenvalue, $\gamma$, of $X^T X$. Thus, we expect the BIA series to converge if $\l < \gamma$. We can place a bound on $\gamma$ using the Gergoshin Circle Theorem \cite{varga2010gervsgorin}. Since the $X_i$ are standardized variables, as the number of samples goes to infinity, $n \rightarrow \infty$,  the matrix element in row $i$ and column $j$ of $X^T X$ is converges to the correlation between $X_i$ and $X_j$,  $r(X_i,X_j)$. Plugging this result into the Gergoshin Cirlcle Theorem gives a bound for the largest eigenvalue, namely  $\gamma \le n(1+ p \tilde{r})$ where $\tilde{r}= \frac{1}{p} \inf_{i} \sum_{j\neq i} |r(X_i, X_j)|$. This suggests that BIA  holds  when $\lambda > \gamma=n(1+ p \tilde{r})$. For large $p$, we can approximate $\tilde{r}$ by the root-mean-squared correlation between features, $r =\sqrt{p^{-1}(p-1)^{-1} \sum_{i \neq j} r^2(X_i, X_j) }$. This defines a natural scale,
\be
\lambda^* = n(1+ p r).
\ee
for the penalty parameter at which the BIA is expected to breakdown. We  expect the BIA to be accurate when $\l \gg \l^*$ and to breakdown when $\l \ll \l^*$.

\subsection{Bayesian Ising Approximation}

Equation \ref{eq:expanded-posterior} still contains design matrices ($X_{\bold{s}}$) that are restricted to the features for which $s_j = +1$. To remove these restrictions, let's introduce some binary indicator variables $\g_j = (s_j+1)/2$. Thus, $\g_j = 1$ if variable $j$ is included in the model and $\g_j = 0$ if variable $j$ is not included in the model. Also, let $V = X^T X/n$ and $G$ be the matrix with elements $G_{ij} = \sum_{k,l= 1}^{k,l = n} y_k y_l X_{ki} X_{lj} / n^2$. Note that standardizing all of the data leads to $V_{ij} = r(x_i,x_j)$ and $G_{ij} \simeq r(y,x_i) r (y,x_j)$, where $r(z_1,z_2)$ is the Pearson correlation between dummy variables $z_1$ and $z_2$.  Now, we will rewrite Eq.\ \ref{eq:expanded-posterior} in terms of the indicator variables ($\boldsymbol{\g}$), $V$, and $G$.  We have:
\begin{align}
\text{Tr}[X_{\bold{s}}^T X_{\bold{s}}] 
& = \sum_{i = 1}^{i = p} \g_i \left( \sum_{j = 1}^{j=n} X_{ji}^2 \right) = n \sum_{i = 1}^{i = p} V_{ii} \g_i  \nonumber
\end{align}
\begin{align}
\text{Tr}[(X_{\bold{s}}^T X_{\bold{s}})^2] 
&= \sum_{i,j =1}^{i,j = p} \g_i \g_j \left( \sum_{k =1}^{k=n} X_{ki} X_{kj}\right)^2 =  n^2 \sum_{i,j =1}^{i,j = p} V_{ij}^2 \g_i \g_j \nonumber
\end{align}
\begin{align}
y^T X_{\bold{s}} X_{\bold{s}}^T y 
&= \sum_{i = 1}^{i = p} \g_i \left(  \sum_{j,k = 1}^{j,k = n} y_j y_k X_{ji} X_{ki} \right) =  n^2 \sum_{i = 1}^{i = p} G_{ii} \g_i \nonumber
\end{align}
\begin{align}
y^T X_{\bold{s}} X_{\bold{s}}^T X_{\bold{s}} X_{\bold{s}}^T y 
& = \sum_{i,j = 1}^{i,j = p} \g_i \g_j \left( \sum_{q = 1}^{q = n} X_{qi} X_{qj} \right) \left( \sum_{k,l= 1}^{k,l = n} y_k y_l 
X_{ki} X_{lj} \right) =  n^3 \sum_{i,j = 1}^{i,j = p} V_{ij} G_{ij} \g_i \g_j  \nonumber 
\end{align}
Plugging these into Eq.\ \ref{eq:expanded-posterior}, we have:
\begin{align}
\mathcal{L}(\bold{s} | y) 
&= \frac{n^2}{2 \l} \left\{ \sum_{i = 1}^{i = p} \left( G_{ii}- \frac{V_{ii} }{n} \right) \g_i  + \frac{n}{\l }  \sum_{i,j =1}^{i,j = p} \left( \frac{V_{ij}^2}{n} -  V_{ij} G_{ij} +\HF G_{ii} G_{jj}  \right) \g_i \g_j \right\} \nonumber
\end{align}
Plugging in $\g_j = (1+s_j)/2$, rearranging and dropping constant terms yields our final result:
\begin{align}
\mathcal{L}(\bold{s} | y)  
&\simeq \frac{n^2}{4 \l}  \left(\sum_{i} h_i(\l) s_i + \HF \sum_{i,j;  i \neq j} J_{ij}(\l) s_i s_j \right)
\end{align}
where the external fields and couplings are defined as:
\begin{align}
h_i(\l) &= r^2(y,x_i) -\frac{1}{n} +\sum_{j} J_{ij}(\l) \\
J_{ij}(\l) &= \frac{n}{\l} \left( \frac{r^2 (x_i,x_j)}{n} -   r(x_i, x_j) r(y,x_i) r(y,x_j) + \HF  r^2(y,x_i) r^2(y,x_j) \right)
\end{align}

\bibliography{refsmain} 

\end{document}